\documentclass[twocolumn]{article}
\usepackage{graphicx} 
\usepackage{tabularx}
\usepackage{rotating}
\usepackage{nameref}
\usepackage{array}
\usepackage{standalone}
\usepackage{comment}
\usepackage{longtable}
\usepackage{makecell}
\usepackage{listings}
\usepackage{xcolor}
\usepackage{caption}
\usepackage{listingsutf8} 
\usepackage{float}       
\usepackage{etoolbox}     
\usepackage{newfloat}     
\usepackage{chngcntr}
\usepackage{authblk}
\usepackage{orcidlink}
\usepackage[labelfont=bf]{caption}
\usepackage[most]{tcolorbox}
\tcbuselibrary{listings, breakable}

\DeclareCaptionType{listing}[Listing][List of Listings]

\usepackage{hyperref} 
\hypersetup{
    colorlinks=true,
    linkcolor=blue,
    filecolor=magenta,      
    urlcolor=blue
    }

\title{The Software Observatory: aggregating and analysing software metadata for trend computation and FAIR assessment}

\date{}

\author[1,2]{Eva Martín del Pico\orcidlink{0000-0001-8324-2897}}
\author[1,2]{Josep Lluís Gelpí\orcidlink{0000-0002-0566-7723}}
\author[1]{Salvador Capella-Gutiérrez\orcidlink{0000-0002-0309-604X}}

\affil[1]{Barcelona Supercomputing Center, Barcelona, Spain}
\affil[2]{Department of Biochemistry and Molecular Biomedicine, University of Barcelona, Barcelona, Spain}

\begin{document}

\maketitle

\begin{abstract}
In the ever-changing realm of research software development, it is crucial for the scientific community to grasp current trends to identify gaps that can potentially hinder scientific progress. The adherence to the FAIR (Findable, Accessible, Interoperable, Reusable) principles can serve as a proxy to understand those trends and provide a mechanism to propose specific actions.

The \textit{Software Observatory} at OpenEBench (\href{https://openebench.bsc.es/observatory}{https://openebench.bsc.es/observatory}) is a novel web portal that consolidates software metadata from various sources, offering comprehensive insights into critical research software aspects. Our platform enables users to analyse trends, identify patterns and advancements within the Life Sciences research software ecosystem, and understand its evolution over time. It also evaluates research software according to FAIR principles for research software, providing scores for different indicators.

Users have the ability to visualise this metadata at different levels of granularity, ranging from the entire software landscape to specific communities to individual software entries through the \textit{FAIRsoft Evaluator}. Indeed, the FAIRsoft Evaluator component streamlines the assessment process, helping developers efficiently evaluate and obtain guidance to improve their software's FAIRness.

The Software Observatory represents a valuable resource for researchers and software developers, as well as stakeholders, promoting better software development practices and adherence to FAIR principles for research software.
\end{abstract}

\section{Introduction}  

Research software is a crucial component of scholarly work, with significant implications for funding, time, and research opportunities \cite{impactstoryteamComprehensiveImpactReport2013}. Its importance is underscored by its widespread use in many data-intensive disciplines, including bioinformatics, as evidenced by its high citation rates and mentions in publications \cite{yangHowImportantScientific2018}. However, challenges in the discoverability, reproducibility, and sustainability of research software persist \cite{barkerIntroducingFAIRPrinciples2022a}. To address these challenges, there is a growing recognition of the need to treat research software as a valuable research output, equivalent to others like research data and peer-reviewed manuscripts \cite{jaySoftwareMustBe2021}. 

In this context, the recently published FAIR principles for research software \cite{chuehongFAIRPrinciplesResearch2022} and their automated implementation using high- and low-level indicators \cite{martindelpicoFAIRsoftPracticalImplementation2024} offer an opportunity to start understanding current practices in research software development, especially those related to software metadata. Software metadata is essential for the findability, accessibility, interoperability, and reusability of research software across scientific domains, including Life Sciences \cite{kuckertzMetadataBasedEcosystemImprove2023}. Machine-readable metadata enhances reproducibility by providing precise and consistent descriptions of software functionalities, dependencies, and versions. This allows researchers to automate its use, either as individual tasks or as part of analytical workflows, ensuring consistent and accurate application of software tools \cite{garijoOKGSoftOpenKnowledge2019}. Additionally, standardised metadata facilitates interoperability between different software systems, reducing errors and enhancing overall efficiency \cite{garijoOKGSoftOpenKnowledge2019}. It also aids in searching, sorting, and analysing software, enabling researchers to quickly assess and compare tools based on objective criteria.

However, the proliferation and distribution of research software information pose significant challenges. Taking as reference the Life Sciences, redundant metadata across multiple repositories and registries leads to inefficiencies and potential confusion regarding the most current or accurate version. Indeed, dispersed information complicates data aggregation efforts with a direct impact on the coherence of available metadata. While centralised registries, like bio.tools \cite{isonBiotoolsRegistrySoftware2019}, where developers may register their software, could address these issues, they may face problems like becoming single points of failure, limited scalability, lack of consistency, and the difficulty of being general enough to meet the diverse and evolving needs of the research community when developing, maintaining, using, and reusing software. Consequently, these challenges hinder the effective analysis and assessment of software FAIRness as a way to understand software quality from a metadata perspective. Advanced tools and methodologies are therefore necessary to automate the collection, standardisation, and analysis of dispersed software metadata for meaningful insights into software FAIRness.

Evaluating the adherence to the FAIR principles of software can be a tedious process, and researchers, who often code themselves, may lack the necessary knowledge and expertise. Efforts have been made to provide user-friendly assessments of software quality. Notable initiatives include ``Self-assessment for FAIR research software'' \cite{spaaksFAIRSoftwareChecklist} and the ``FAIR-Checker'' \cite{gaignardFAIRCheckerSupportingDigital2023}, both of which evaluate quality in terms of Findability, Accessibility, Interoperability, and Reusability. Additionally, the ``Open Source Security Foundation (OpenSSF) Best Practices Badge'' \cite{BadgeApp} provides a comprehensive set of good practices categorised by levels of maturity. These initiatives underscore the importance of systematic and user-friendly approaches to improving software quality and adherence to FAIR principles.

As a practical implementation of the FAIRsoft indicators \cite{martindelpicoFAIRsoftPracticalImplementation2024}, the Software Observatory at OpenEBench\footnote{\url{https://openebench.bsc.es/observatory}} offers a comprehensive and scalable platform for evaluating the FAIRness of Life Sciences research software. By aggregating metadata from diverse sources—including bio.tools, Bioconda\cite{gruningBiocondaSustainableComprehensive2018}, Bioconductor\cite{gentlemanBioconductorOpenSoftware2004}, and others—the Observatory not only consolidates dispersed records through disambiguation and enrichment, but also enables large-scale, community-level analysis. It distinguishes itself from isolated checkers by integrating automated FAIR assessments with interactive dashboards that support exploration of temporal trends, domain-specific benchmarks, and project-level insights. Through the FAIRsoft Evaluator, it further provides actionable guidance for improving metadata quality and promoting good practices. Designed to support researchers, software developers, curators, and policy-makers, the Observatory facilitates informed decisions and progress tracking across the research software landscape.

\section{Materials and Methods}

\subsection{Metadata extraction, normalisation, and enrichment}

\begin{figure*}[h!]
    \centering
    \includegraphics[width=1\linewidth]{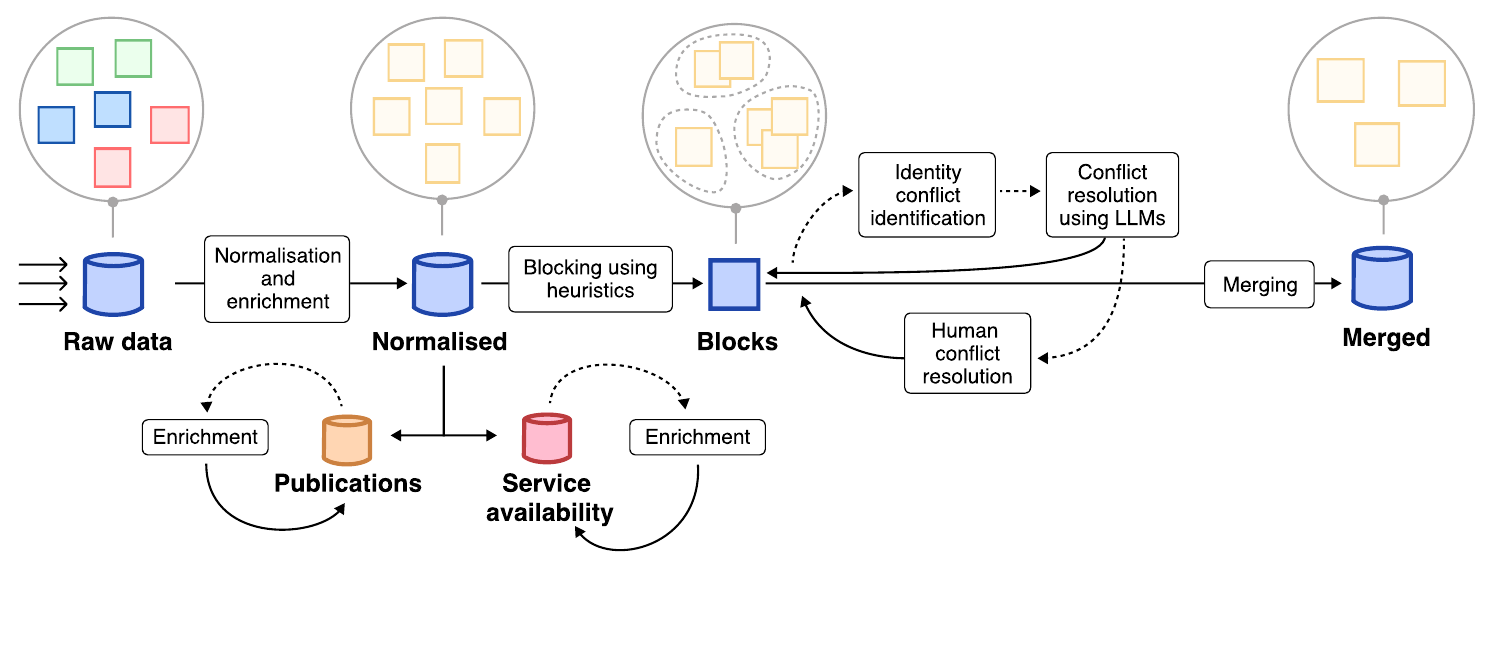}
    \caption{\textbf{Software Observatory metadata processing pipeline}. Metadata is ingested from external sources into a raw collection, enriched and normalized through automated methods, and integrated into a deduplicated final dataset. Internal enrichment includes SPDX license mapping, EDAM format normalization, and contributor classification, while auxiliary metadata (e.g., publication data from Europe PMC and Semantic Scholar, and service availability) is retrieved via decoupled pipelines. The integration step involved grouping records into blocks, identifying potential conflicts within each block, and resolving them through a combination of heuristic rules, LLM-based assessments, and human validation. The block structure used for conflict resolution is stored in a dedicated, persistent state file, which is updated after each resolution step and acts as the source of truth for the merged collection. Once resolved, records within each block were merged into unified entries, resulting in an independent, deduplicated collection that constitutes the final output of the pipeline.\\
    This layered architecture separates ingestion, enrichment, and integration stages, enabling independent evolution of each component. Intermediate layers store enriched-but-unmerged entries to support traceable, incremental updates without re-importing the full dataset. Manual disambiguation and logic improvements (e.g., parsing heuristics) can be applied at the enrichment stage, enhancing flexibility, reproducibility, and adaptation to evolving metadata standards. The block file, maintained as a persistent source of truth, captures the grouping logic used in integration and ensures consistency between conflict resolution and the merged collection. Its persistence allows for transparent correction, re-evaluation, and downstream reproducibility}
    \label{fig:pipeline}
\end{figure*}

The Software Observatory processes software metadata through a modular pipeline designed to consolidate, enrich and standardise information from diverse sources (Figure~\ref{fig:pipeline}). Metadata is initially ingested from registries such as bio.tools, Bioconda, Bioconductor, Galaxy ToolShed\cite{blankenbergDisseminationScientificSoftware2014}, SourceForge\cite{SourceForge} (bioinformatics-tagged), and Galaxy Europe\cite{afganGalaxyPlatformAccessible2016}. In addition, GitHub\cite{GitHub} repositories are mined by following links present in these primary records, allowing the extraction of further metadata such as license, README, and contributors. This raw metadata forms the basis for a structured integration workflow composed of two main branches: internal normalisation and external enrichment. Further details on the implementation of this process, including the specific components used at each stage, are provided in Tables S1 and S2.

Internally, fields such as input/output formats and licence information are harmonised to ensure consistency across datasets. Format types are cleaned and mapped to EDAM~\cite{isonEDAMOntologyBioinformatics2013} terms, while licence information is matched to SPDX License List\cite{SPDXLicenseList} identifiers using a curated synonym list\cite{evamartindelpicoLicensesmappingToolMap}. Early experiments with similarity-based string matching proved unreliable, so curated mappings were preferred for their higher precision and lower false-positive rate.

In parallel, auxiliary metadata is retrieved from external services to enhance the completeness and usability of each record. This includes publication metadata from Europe PMC\cite{theeuropepmcconsortiumEuropePMCFulltext2015} and Semantic Scholar\cite{kinneySemanticScholarOpen2023} (e.g., citation counts, abstract, publication year), as well as service availability metrics obtained through direct checks on the tools’ webpages for those classified as deployable services (e.g., web, REST, SPARQL). While these enrichments are performed separately from the core integration pipeline, they are ultimately unified in the final dataset to support comprehensive FAIR assessment.

Throughout the process, cleansing operations—such as pruning malformed fields, deduplicating links, and applying consistent formatting—are applied to improve machine-readability and reduce integration errors. The resulting enriched and normalised records are stored in a dedicated intermediate database ("Normalised" in Figure \ref{fig:pipeline}), serving as the foundation for subsequent stages like identity resolution and FAIR scoring.

\subsection{Metadata integration and disambiguation}

Integrating metadata from heterogeneous registries requires resolving cases where the same software is referenced under different names, formats, or incomplete links. We implemented a multi-stage disambiguation pipeline designed to detect and resolve such cases while minimizing erroneous merges.

\subparagraph{Conservative grouping via name and link similarity}

As a first step, we grouped software records into candidate blocks using a conservative blocking strategy. Entries were grouped if they shared either (i) a normalized software name and type, or (ii) at least one normalized repository or recognized repository-like webpage link (e.g., GitHub, Bioconductor, SourceForge). Link normalization included domain-specific rules to collapse variants of URLs and ensure transitive linkage across records. This approach avoids overmerging by requiring structural evidence—such as a shared repository—in addition to name similarity.

\subparagraph{Conflict detection within grouped blocks}

Within each group, we identified identity conflicts as cases where records were grouped by name but lacked any shared repository or repository-like link. These cases present a high risk of incorrect merging, as they typically involve unrelated tools that happen to share a name.

Other forms of ambiguity—such as records with different names but shared repositories or closely related functionality—are also important but were not addressed in this initial disambiguation benchmark. These often involve tools that are conceptually or functionally linked (e.g., plugins, forks, or dependencies), and may require more nuanced resolution strategies than name-based cases.

\subparagraph{Conflict resolution via rescue heuristics and Large Language Models (LLMs)}

To resolve the targeted conflicts, we first applied a rescue heuristic to reduce false positives in the conflict set. Specifically, if a record initially flagged as disconnected shared both a name and source (e.g., the same registry) with an accepted group member, it was promoted into the group. This allowed us to recover plausible matches that lacked repository links but were likely to refer to the same software tool  based on consistent naming and shared metadata fields. For instance, several \texttt{gromacs} records (types cmd, lib, suite) sharing the webpage "www.gromacs.org" were merged, while \texttt{gromacs\_mpi} was excluded due to its divergent name. Similarly, \texttt{anvio} records of types workflow and cmd were grouped via a shared URL (merenlab.org/software/anvio), but \texttt{anvio-minimal} was left out. When two records share both a name and a webpage, it is unlikely they refer to different tools—the name effectively acts as a local identifier. In contrast, when only the webpage is shared and names diverge, the records may simply reflect an association, such as tools from the same laboratory, project or family, rather than identity.
The resulting, refined conflict set was then passed to the resolution phase. 

We addressed the remaining conflicts using a hybrid approach combining automation and human review. A large language model-based agreement proxy~\cite{martindelpicoIdentityResolutionSoftware2025} evaluated each conflict block by comparing the semantic similarity of metadata fields, README content, and associated webpages, and successfully resolved a substantial portion of cases. Ambiguous blocks were escalated to human reviewers via structured GitHub issues, ensuring transparent decision-making and integration through GitHub Actions.

All decisions—automated or manual—were stored persistently, forming a growing annotated dataset for future refinement and model retraining. After conflict resolution, all records within a block were merged to produce the final integrated set of software metadata entries—represented as the “Merged” layer in Figure~\ref{fig:pipeline}. This merged dataset serves as the foundation for all visualizations provided by the Software Observatory and also acts as one of the metadata sources used by the FAIRsoft Evaluator. FAIRness scores, completeness indicators, and aggregation statistics are precomputed from this final collection and stored in a dedicated database to support efficient rendering in the web interface.

\subsection{User interface and functionalities}

\begin{figure*}[h!]
    \centering
    \includegraphics[width=0.9\textwidth]{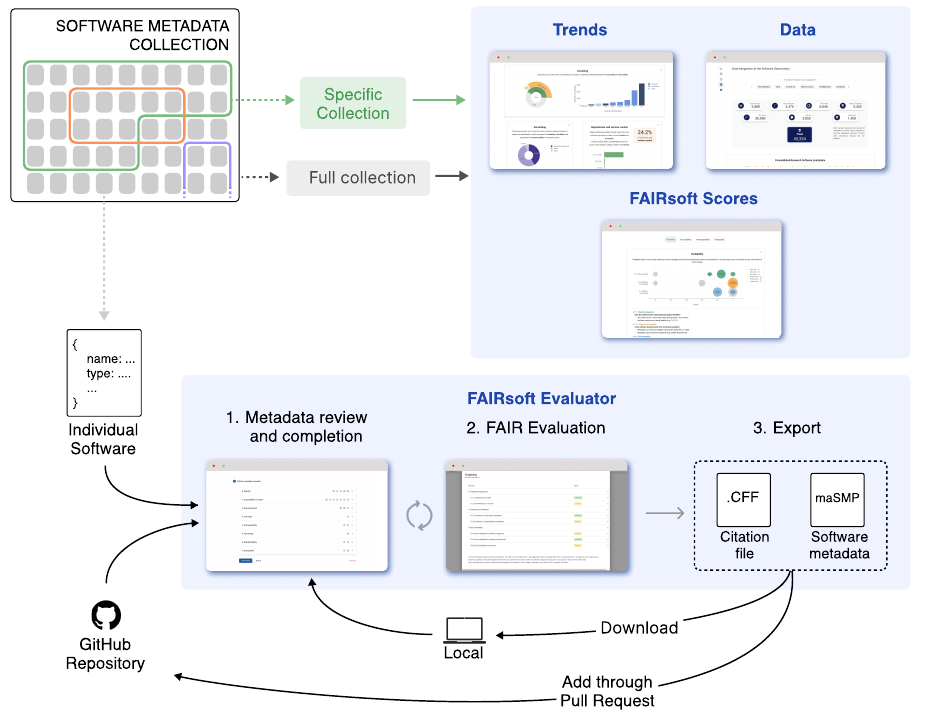}
    \caption{\textbf{Functional components of the Software Observatory user interface.} The platform supports exploration of the full software collection, specific communities or projects, and individual entries. Visual dashboards display trends, data aggregation details, and FAIRsoft scores. Individual software entries can be evaluated through the FAIRsoft Evaluator in three steps: (1) metadata review and completion, (2) FAIR assessment, and (3) export of metadata in formats such as \texttt{.CFF} (citation file) and \texttt{maSMP} (a structured JSON-LD profile compatible with CodeMeta, Bioschemas, and schema.org). The tool supports both local downloads and direct GitHub integration via pull requests, facilitating metadata improvement and reuse.}
    \label{fig:observatory_functionalities}
\end{figure*}

The Software Observatory features a modular web-based interface designed to support both high-level metadata exploration and detailed FAIRness assessment. As illustrated in Figure~\ref{fig:observatory_functionalities}, it is structured into four main components:

\begin{itemize}
    \item \textbf{Trends}: Provides visual summaries of metadata attributes such as licensing, versioning, and repository usage. These charts enable users to track community practices and longitudinal changes.
    
    \item \textbf{FAIR Scoreboard}: Displays aggregated FAIRsoft indicator scores across the dataset. Users can filter by project or community and embed individual charts via iframe integration, facilitating dissemination on external websites.
    
    \item \textbf{Data}: Offers detailed statistics on metadata completeness, source contributions, and integration coverage. Filters allow users to focus on specific communities or domains.
    
    \item \textbf{FAIRsoft Evaluator}: Enables software-level FAIRness assessment. It supports metadata retrieval from various sources (e.g., GitHub, local files, and the Observatory database), provides guided editing, calculates FAIRsoft scores, and facilitates metadata export in the maSMP format\cite{olgaMachineactionableSoftwareManagement2023}—a structured profile that integrates entities from standards such as CodeMeta\cite{CodeMetaProject}, Bioschemas\cite{grayPotatoSaladProteinb}, and schema.org\cite{Schemaorg} to ensure broad compatibility. It also supports direct integration with GitHub via automated pull requests.
\end{itemize}

Together, these components empower both individual software developers and community leads to evaluate and improve research software metadata. The interface promotes FAIR-aware practices and provides actionable guidance for enhancing software quality.

\vspace{1em}

\section{Results}

\subsection{Metadata aggregation and enrichment}

\begin{figure*}[h!]
    \centering
    \includegraphics[width=1\linewidth]{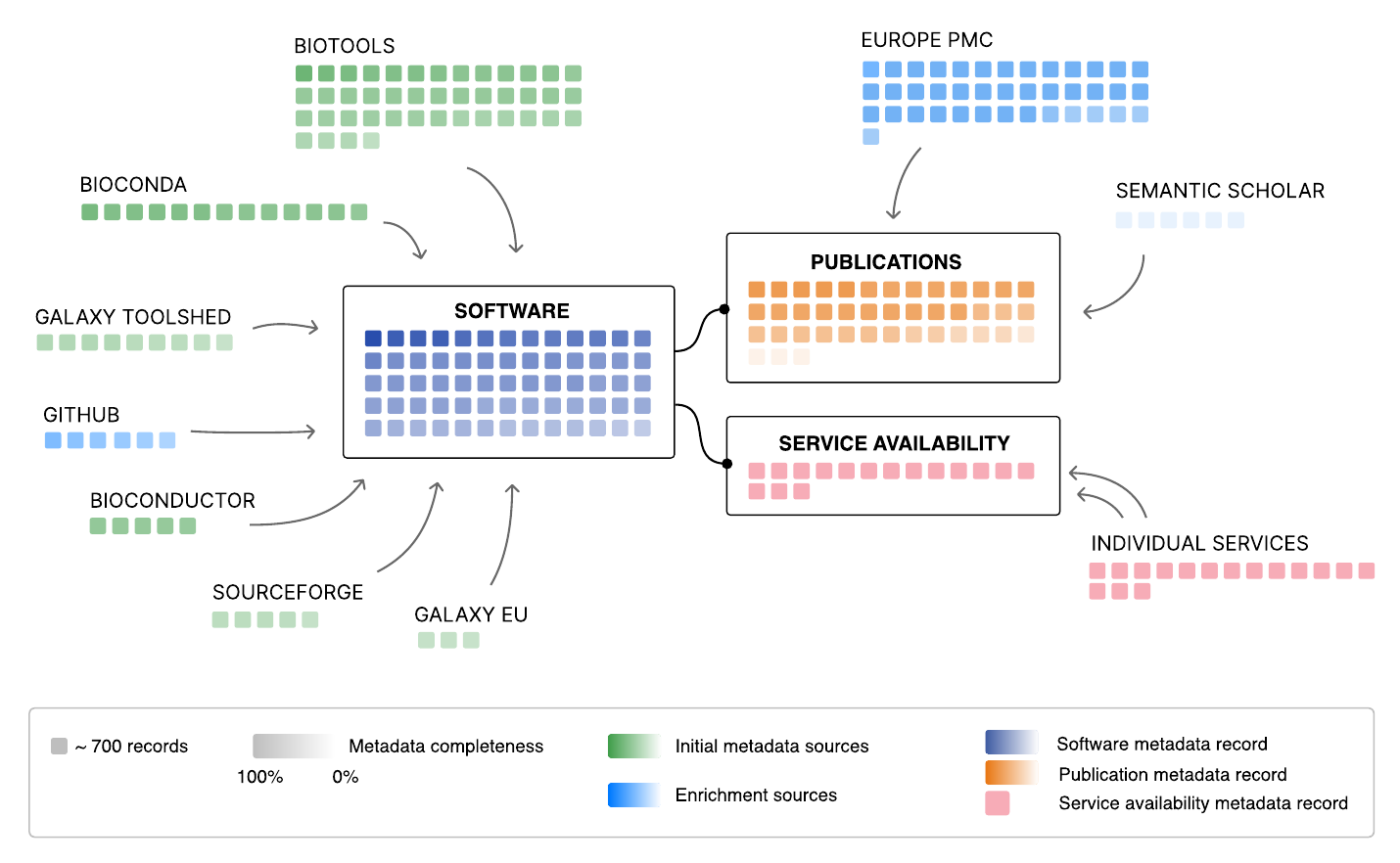}
    \caption{\textbf{Sources of metadata integrated into the Software Observatory.} Observatory aggregates software metadata from a range of initial metadata sources (green), including registries such as bio.tools, Bioconda, the Galaxy ToolShed, Bioconductor, SourceForge, Galaxy Europe, and linked GitHub repositories. These records are subsequently enriched using external enrichment sources (light blue), including Semantic Scholar and Europe PMC for publication metadata, and direct service availability checks for deployable tools. These enrichment processes rely on identifiers (e.g., DOIs, service URLs, and software type) already present in the collected software metadata. This figure shows the contribution of each external source to the final integrated dataset. Each square corresponds to approximately 700 software metadata records. The structure of the figure mirrors the layered, dependency-aware integration architecture outlined in Figure ~\ref{fig:pipeline}.}
    \label{fig:sources}
\end{figure*}
s

The Software Observatory initially collected over 65,942 metadata records from major Life Sciences registries, including bio.tools, Bioconda, and the Galaxy ToolShed. Many of these records referred to the same software tools, resulting in redundancy and overlap across the dataset. After normalization, disambiguation, and integration, the final dataset comprised 45,334 unique software records, which form the basis for all enrichment statistics and downstream analyses reported in this work.

The integrated dataset aggregates contributions from several primary sources. The largest contributor is bio.tools (30,063 records), followed by Bioconda (8,846), the Galaxy ToolShed (6,302), and Bioconductor (3,470). Other relevant sources include SourceForge (2,989) and Galaxy Europe (1,409). Additional metadata was extracted from GitHub (3,852), which was mined following links present in the previously collected records.

Metadata processing includes both normalization steps and auxiliary enrichments. Core normalization—conducted within the software-centric integration pipeline—standardized license information using SPDX identifiers (7,005 normalized declarations, covering 34.6\% of license-bearing records), mapped software formats to EDAM ontology terms (26,464 values across 3,931 records), and heuristically classified authors as either persons or organizations. At present, only a single "author" role is tracked, often conflating maintainers and developers—a known limitation and area for future refinement.

Other enrichments were performed through decoupled systems designed around non-software-specific data types. Availability of service was assessed for 11,339 deployable tools (e.g., \textit{web}, \textit{REST}, \textit{SPARQL}, \textit{SOAP}, \textit{workbench}, or \textit{suite}) using real-time URL responsiveness checks. Publication metadata was retrieved from Europe PMC for 28,271 publications with total citation counts. Citations per year were extracted from Semantic Scholar for 4,403 publications. These enrichments are maintained in auxiliary databases and are not part of the integration pipeline shown in Figure~\ref{fig:sources}.

\subsection{Disambiguation of software identities}

\begin{figure}[h!]
    \centering
    \includegraphics[width=1\linewidth]{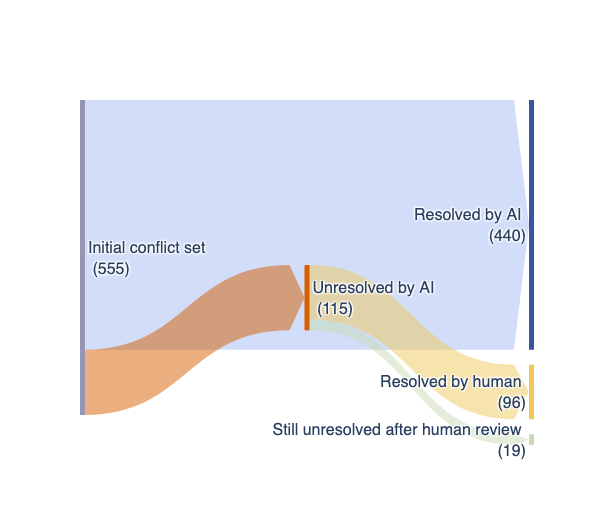}
    \caption{Disambiguation pipeline and conflict resolution outcomes. Top: metadata entries are grouped and flagged as potential conflicts, which are resolved using a hybrid approach of LLM-based agreement proxies and human curation. Bottom: distribution of entries, showing proportions resolved automatically, escalated, or discarded.}
    \label{fig:disambiguation-workflow}
\end{figure}

A total of 555 blocks were identified as containing identity conflicts, representing approximately 1.2\% of all grouped software metadata records. After applying a rescue heuristic to exclude likely matches based on shared name and source, 440 conflict blocks remained and were resolved automatically using a large language model-based agreement proxy. This mechanism selected the most semantically consistent resolution for each block, automating roughly 79\% of all final integration decisions. The remaining cases, flagged as ambiguous by the model, were reviewed and resolved manually through structured GitHub issues, ensuring both transparency and reproducibility.

All decisions were recorded in a persistent conflict registry, supporting traceability, auditability, and future refinement as metadata sources evolve. The resulting disambiguated dataset underlies both the FAIRness scoring pipeline and the interactive visualizations available through the Software Observatory interface. A schematic overview of the full disambiguation workflow is shown in Figure~\ref{fig:disambiguation-workflow}.

\subsection{Case study: the Proteomics community}

\begin{figure*}[ht]
  \centering
  \includegraphics[width=1\textwidth]{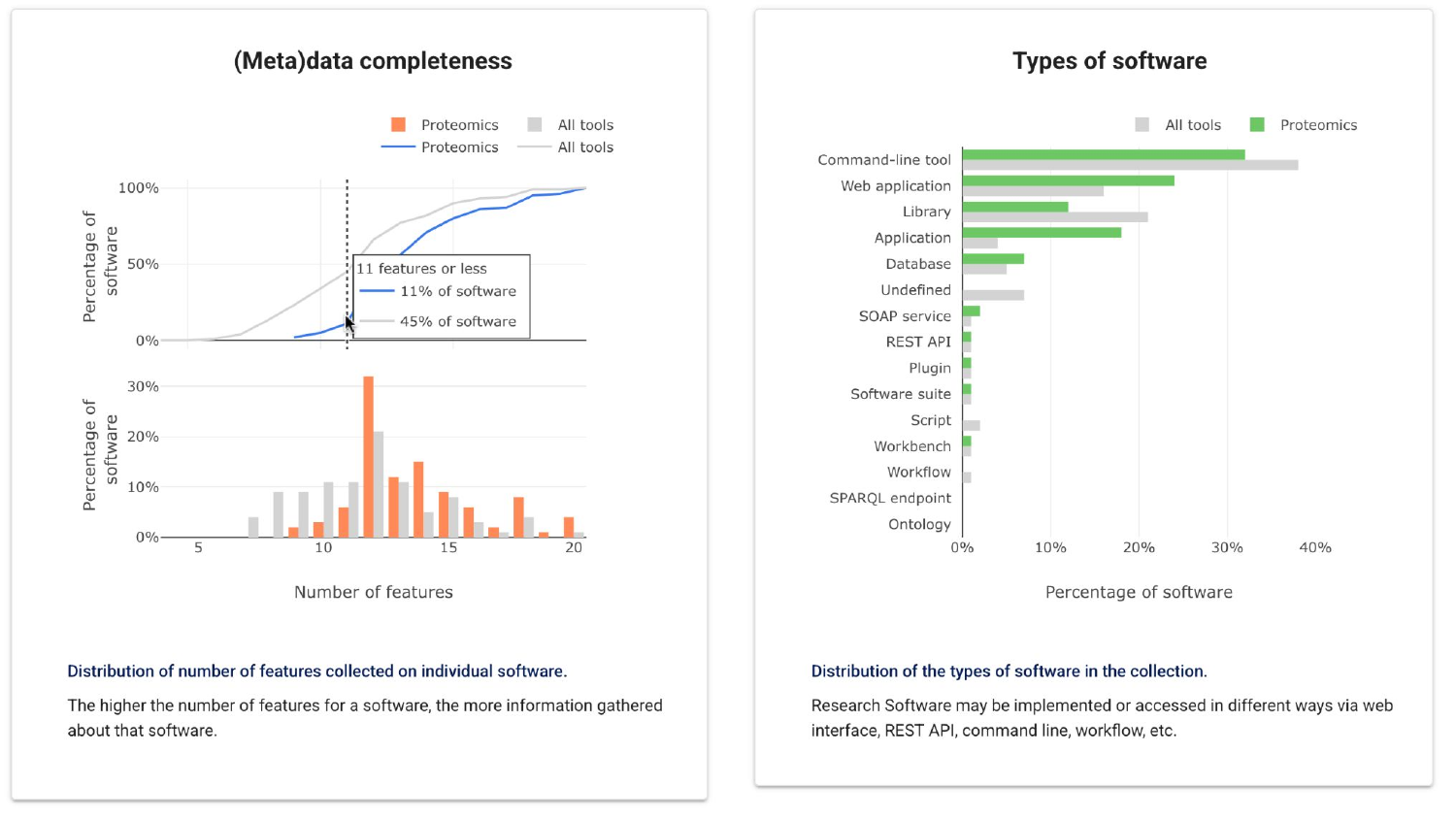}
  \caption{\textbf{Interactive visualization of (meta)data completeness and software types for the Proteomics community in the Software Observatory.}}
  \label{fig:proteomics-completeness}
\end{figure*}

To demonstrate the Observatory’s capacity for targeted community-level analysis, we examined the software ecosystem affiliated with the ELIXR \textit{Proteomics} project. This collection comprises 758 tools explicitly linked to the project through curated metadata in \textit{bio.tools}, including the project identifier, contributor details, software descriptions, and publication among others. Figure~\ref{fig:proteomics-completeness} illustrates how the Observatory interface enables interactive exploration of metadata completeness and software type distributions for this collection. 

Figures S6–S9 present the FAIRsoft Scoreboard for the ELIXIR Proteomics community, providing a detailed overview of indicator-level performance across the four FAIR principles. The Data interface, which offers insight into metadata source provenance and coverage, is illustrated in Figures ~\ref{fig:proteomics-completeness}, S10 and S11. Figure S2 shows how the collection is introduced within the Observatory interface, while Figures S3–S5 highlight specific panels from the FAIRness Trends Analysis interface. Finally, Figures~S12–S16 detail the main steps of the FAIRsoft Evaluator, including metadata source selection, refinement, evaluation, and the interpretation of individual indicator results.

Figure S10 presents the provenance of these records. Most tools come exclusively from bio.tools (611), with a minority of records extracted from other sources, such as   Bioconda (100), Bioconductor (86) or GitHub (28). 

Metadata quality in this community is generally high. Essential fields, such as name, type, description, webpage, topics and author, were universally reported. Several others, including publication links (89\%), documentation (86\%), also showed strong coverage. Nonetheless, important elements for accessibility were frequently absent. For example, only 10\% of tools reported testing information or explicit dependency declarations and only 20\% provided download links. It is important to note that the bio.tools registry does not currently support metadata fields for testing or dependencies. Therefore, these findings may reflect limitations in the metadata schema rather than deficiencies in development practices, that have an impact in the FAIRsoft scores.

FAIRsoft Evaluator results reveal a nuanced FAIRness profile:
\begin{itemize}
    \item Findability (F) emerged as a key strength in the Proteomics community. Metadata registration was exceptionally high, with 100\% of tools achieving a perfect score on indicator F2, reflecting the consistent generation of rich metadata annotated with the EDAM ontology. Discoverability (F3) was also notable: the majority of tools were linked to a publication, and their presence in bio.tools further enhanced their visibility within the scientific literature. Moreover, Proteomics tools appeared in leading journals, including Nucleic Acids Research, Bioinformatics, and Nature Methods. These publications collectively accounted for hundreds of citations, with Nature Methods alone contributing over 600. Figure S5 summarizes the distribution of citations across publication venues.
	\item Accessibility (A) remains limited. Indicator A1, “Availability of a working version,” had a low mean score of 0.14, primarily due to the absence of explicit download links and installation documentation. This issue is particularly critical given that the collection consists mainly of command-line tools and libraries. Additionally, the limited availability of these tools in e-infrastructures such as Galaxy further contributes to their reduced accessibility.
	\item Interoperability (I) was moderate. While support for standard data formats (I1) was above average and generally well documented, integration with other software (I2) was weak. This is largely due to the low proportion of tools offering programmatic interfaces—such as libraries or APIs—and to the community’s limited representation in platforms like Galaxy, which are key for facilitating interoperable workflows.
	\item Reusability (R) showed a mixed pattern. The community performed strongly in contribution recognition, with 100\% of tools reporting authorship (R3). Usage documentation was also more frequently available than average, although the improvement was modest. In contrast, licensing (R2) remains a major weakness: 70\% of tools lacked any declared license, significantly limiting clarity on reuse conditions. Among the 229 tools (30\%) with unambiguous licensing information, the majority (203) were open source. Copyleft licenses such as GPL were most common (102 tools), followed by permissive licenses like Apache (41), MIT (18), and Artistic (22). Finally, versioning and version control (R4) were average: only 19\% of tools declared the use of version control systems (e.g., Git), and 30\% provided evidence of software versioning.

\end{itemize}

Taken together, these findings highlight the strengths of the ELIXIR Proteomics software community while also exposing gaps in engineering practices and interoperability. The Software Observatory makes these detailed assessments possible, offering interactive visualizations and metric breakdowns to support community-driven improvements. All metrics and dashboards for the Proteomics collection are accessible at \url{https://openebench.bsc.es/observatory}.

\section{Discussion}

 The Software Observatory addresses a critical challenge in the Life Sciences and beyond the fragmentation, redundancy and inconsistency of software metadata across registries and repositories. By aggregating, normalizing and disambiguating metadata from multiple sources, the Sofware Observatory enables large scale, automated asessment of software FAIRness. In this section, we interpret our main findings in the context of existing efforts, highlight the strengths and limitations of our approach, and outline implications for software developers, funders, and the broader research community.

\subsection{Insights from the disambiguation process}
A major challenge encountered during the disambiguation process was the frequent presence of broken or outdated URLs, which impeded both manual inspection and automated verification. Among the deployable services examined, approximately 44\% of URLs were found to be unavailable. This high failure rate highlights a broader structural issue: the research software ecosystem remains heavily dependent on transient web resources. This reliance undermines long-term discoverability, complicates identity resolution, and poses a serious threat to the sustainability of research software. Similar patterns have been documented for research data, where availability declines significantly over time, largely due to the absence of robust archival practices and the diminishing accessibility of corresponding authors \cite{vinesAvailabilityResearchData2014}. These parallels underscore the urgent need for systemic policies and infrastructures that ensure the persistent accessibility of both research software and data.

\subsection{Lessons from the ELIXIR Proteomics community}

We selected the Proteomics community as a case study due to its well-established engagement with structured software registration, particularly via platforms like \textit{bio.tools}. Our analysis confirmed the community’s strong commitment to metadata quality: the vast majority of tools include rich descriptive metadata, publication references, and documentation, resulting in high performance on findability and reusability indicators. Universal assignment of persistent identifiers and widespread license declarations underscore this strength.

Our analysis of the Proteomics community revealed a notable absence of crucial metadata—such as information on testing and dependencies—that resulted in low scores for accessibility and reusability. However, this finding requires careful interpretation. These gaps may not necessarily reflect poor software development practices, but rather limitations in the metadata sources themselves. In this case, the majority of tools were drawn exclusively from bio.tools, a registry designed primarily for classification and discoverability rather than for capturing development-related metadata. For instance, bio.tools does not offer dedicated fields for reporting testing infrastructure or software dependencies, meaning that even well-tested software may appear deficient in this regard. In contrast, platforms like the Galaxy ToolShed do support such metadata, but are underrepresented in this dataset.

This highlights the importance of considering the scope and purpose of each registry when interpreting metadata-driven evaluations. No single source captures the full spectrum of information required for comprehensive software assessment. Therefore, aggregating metadata from diverse platforms is essential for constructing a more accurate and nuanced picture of research software quality, sustainability, and FAIRness.

\subsection{FAIRsoft Evaluator: supporting improvement workflows}

The FAIRsoft Evaluator helps mitigate these limitations by providing an interactive environment where users can manually supply missing information—such as the presence of testing infrastructure, access restrictions (e.g., registration requirements), and other relevant evidence not available in the original metadata. This makes the Evaluator a hybrid tool: automated and scalable, yet adaptable enough to incorporate expert knowledge and user-provided context.

The FAIRsoft Evaluator is designed not only to assess FAIRness but to guide metadata improvement. Evaluating adherence to FAIR principles can be tedious and technically demanding—especially for developers who are primarily researchers. The Evaluator lowers this barrier by providing a developer-centric experience that aligns with principles of simplicity, guidance, and actionable feedback. This guidance allow developers to prioritize improvements according to their project context. Not all indicators are equally relevant to every tool, and the Evaluator supports informed, context-aware decision making. Once metadata has been edited, users can export it in standardized formats  such as maSMP (Bioschemas, CodeMeta, and schema.org-compliant JSON-LD), or create pull requests to update GitHub repositories directly.

\subsection{Limitations and future directions}

Despite the robustness of the aggregation and normalization pipeline, some key metadata domains remain only partially addressed. For instance, author information is currently stored as raw, unintegrated fields, preventing meaningful analysis of collaboration networks across software projects. However, in the absence of consistent use of persistent identifiers such as ORCIDs, common issues such as duplicated authors, varying email addresses, and slight name differences remain unresolved, limiting the ability to trace contributor activity over time or across domains. Similarly, while the current pipeline relies on structured metadata from upstream sources, it does not yet incorporate metadata mining from associated documents such as README files, LICENSE texts, or configuration files. These sources could provide valuable information on usage, dependencies, system requirements, or institutional affiliations, particularly for projects with sparse formal metadata. Integrating author disambiguation and document-based metadata extraction are active areas for future development that could significantly enrich the Observatory’s analytical capabilities. 

Currently, the system handles one class of conflict—records that share a name but lack a common repository—affecting approximately 1.2\% of blocks. Our next goal is to extend disambiguation strategies to more prevalent conflict patterns, such as cases where records differ in name but share a repository. These refinements will help improve recall and robustness while continuing to support transparency and human oversight. 

An additional limitation concerns the way FAIRsoft indicators are currently visualized. The existing interface presents all indicators with equal visual weight, which may inadvertently suggest that they contribute equally to a tool’s overall FAIRness score. However, in the underlying FAIRsoft model, indicators are weighted differently when computing the total score for each principle. For example, interoperability indicator I2—assessing integration with other software—has a weight of 0.1, while I1 and dependency availability carry weights of 0.6 and 0.3, respectively. As a result, weak performance on I2 has a relatively minor impact, yet this nuance is not conveyed in the current display. Future improvements should focus on enhancing the visual encoding of these weights and clarifying their influence on scoring to support more accurate interpretation and prioritization by users.


\section{Conclusions}

The Software Observatory addresses a critical challenge in the life sciences and beyond: the fragmentation, redundancy, and inconsistency of software metadata across registries and repositories. By aggregating, normalizing, and disambiguating metadata from multiple sources, the Observatory enables large-scale, automated assessment of software FAIRness. Its modular pipeline supports traceable integration, while the FAIRsoft Evaluator component empowers developers to engage directly with FAIR principles through guided, editable assessments and standard-compliant metadata export.

Through a case study of the Proteomics community, we demonstrated the Observatory's ability to surface both best practices and persistent gaps in software metadata quality. Even in a domain with strong documentation and registration habits, limitations in accessibility, archival stability, and structured interoperability were made visible. These insights highlight the Observatory's value not only as an assessment platform but also as a feedback mechanism to guide community improvement.

The Observatory's web interface offers continuously updated dashboards and flexible visualizations for researchers, developers, and community coordinators. Early adoption of its FAIRness evaluation tools within live repositories shows promise for broader integration into software development workflows.

Looking ahead, planned enhancements include author disambiguation, metadata mining from software documentation (e.g., README files), and support for longitudinal tracking. With these developments, the Software Observatory will continue to evolve as a strategic infrastructure for understanding and improving the sustainability, accessibility, and impact of research software.

\bibliographystyle{plainurl} 
\bibliography{SoftwareObservatory}

\clearpage

\end{document}